\documentclass[a4paper,12pt]{article}

\usepackage{amsmath}
\usepackage{amssymb}
\usepackage{graphicx}
\begin{document}
\def\textindent#1{\indent\llap{#1\enspace}\ignorespaces}
\renewcommand{\baselinestretch}{1.3}\small\normalsize

\title{\bf Nonmonotonic size dependence of the critical concentration in 2D percolation of straight rigid rods under equilibrium conditions} \vspace{1mm}

\renewcommand{\baselinestretch}{1.0}\small\normalsize
\author{D. A. Matoz-Fernandez, D. H. Linares and A. J. Ramirez-Pastor{\thanks{
\underbar{\bf Corresponding author}: Dr. A. J. Ramirez-Pastor,
Departamento de F\'{\i}sica, Instituto de F\'{\i}sica
Aplicada, Universidad Nacional de San Luis-CONICET, Chacabuco 917,
D5700BWS San Luis, Argentina
{\bf Email:~} antorami@unsl.edu.ar} }\\[5mm]
Departamento de F\'{\i}sica, Instituto de F\'{\i}sica Aplicada, \\
Universidad Nacional de San Luis-CONICET, \\
Chacabuco 917, D5700BWS San Luis, Argentina
}
\maketitle
\vspace{0cm}
\begin{abstract}
Numerical simulations and finite-size scaling analysis have been
carried out to study the percolation behavior of straight rigid
rods of length $k$ ($k$-mers) on two-dimensional square lattices.
The $k$-mers, containing $k$ identical units (each one occupying a
lattice site), were adsorbed at equilibrium on the lattice. The
process was monitored by following the probability
$R_{L,k}(\theta)$ that a lattice composed of $L \times L$ sites
percolates at a concentration $\theta$ of sites occupied by
particles of size $k$. A nonmonotonic size dependence was observed
for the percolation threshold, which decreases for small particles
sizes, goes through a minimum, and finally asymptotically
converges towards a definite value for large segments. This
striking behavior has been interpreted as a consequence of the
isotropic-nematic phase transition occurring in the system for
large values of $k$. Finally, the universality class of the model
was found to be the same as for the random percolation model.
\end{abstract}
\renewcommand{\baselinestretch}{1.3}\small\normalsize
\vspace{1mm}
\vspace{1mm}
\renewcommand{\baselinestretch}{1.3}\small\normalsize
\newpage

\section{Introduction} \label{Introduction}

The percolation problem is a topic being increasingly considered
in statistical physics. One reason for this current interest is
that it is becoming clear that generalizations of the pure
percolation problem are likely to have extensive applications in
science and technology
\cite{Zallen,Stauffer,Sahimi,Grimmett,Bollo}. Although it is a
purely geometric phenomenon, the phase transition involved in the
process can be described in terms of an usual second-order phase
transition. This mapping to critical phenomena made percolation a
full part of the theoretical framework of collective phenomena and
statistical physics.

Despite of the number of contributions to this problem, there are
many aspects which are not yet completely solved. In fact, most of
the studies are devoted to the percolation of objects whose size
coincides with the size of the lattice site (single occupancy).
However, if some sort of correlation exists, like particles
occupying several $k$ contiguous lattice sites ($k$-mers), the
statistical problem becomes exceedingly difficult and a few
studies have been devoted to understanding the percolation of
elements occupying more than one site (bond)
\cite{Harder,Gao,Holloway,Dolz,Leroyer,Bonnier,Cornette,EJPB1,Cornette2,Cornette3,Cherkasova,LONGO}.

In two previous articles \cite{EJPB1,LONGO}, referred to as papers
I and II, respectively, the percolation of straight rigid rods on
2D lattices was studied. In paper I, linear $k$-mers were
deposited randomly and irreversibly. It was established that (1)
the percolation threshold exhibits a exponentially decreasing
function when it is plotted as a function of the $k$-mer size and
(2) the problem belongs to the same universality class that the
random percolation regardless the value of $k$ used. Paper II was
devoted to the study of aligned rigid rods on square lattices. In
this case, the $k$-mers were irreversibly deposited along one of
the directions of the lattice. The results obtained revealed that
(i) the percolation threshold monotonically decreases with
increasing $k$; (ii) for any value of $k$ ($k>1$), the percolation
threshold is higher for aligned rods than for isotropically
deposited rods; (iii) the phase transition occurring in the system
belongs to the standard random percolation universality class; and
(iv) the intersection point of the percolation cumulant curves for different system sizes exhibits a nonuniversal
critical behavior, varying continuously with changing the $k$-mer
size.

On the other hand, numerous experimental and numerical studies
have been recently devoted to the analysis of equilibrium
properties in systems of non-spherical particles
\cite{VIAMONTES,DEMICHE,VINK,CUETOS,GHOSH,EPL1,PHYSA19,JCP7,JSTAT1}.
Of special interest are those studies dealing with lattice
versions of the present problem, where the situation is much less clear
than in the continuum case. In this line, a system of straight
rigid rods of length $k$ on a square lattice, with two allowed
orientations, was recently studied by Monte Carlo simulations
\cite{GHOSH}. The authors found strong numerical evidence on the
existence of an isotropic-nematic (IN) phase transition at
intermediate densities for $k \geq 7$. The nematic phase,
characterized by a big domain of parallel $k$-mers, is separated
from the isotropic phase by a continuous transition occurring at a
finite density. The accurate determination of the critical
exponents, along with the behavior of Binder cumulants, showed
that the transition from the low-density disordered phase to the
intermediate-density ordered phase belongs to the 2D Ising
universality class for square lattices and the three-state Potts
universality class for honeycomb and triangular lattices
\cite{EPL1,PHYSA19}. In addition, the comparison between the
configurational entropy of the system and that corresponding to a
fully aligned system \cite{JCP7,JSTAT1} (i) confirmed the
existence of a IN phase transition at intermediate densities for
long rods, (ii) allowed us to estimate the minimum value of $k$,
which leads to the formation of a nematic phase, and (iii)
provided an interesting interpretation of this critical value.

Later, the dependence of the critical density on the magnitude of
the lateral interactions was studied for a system of attractive
rigid rods on square lattices with two allowed orientations
\cite{PIP1,JCP11}. The obtained results revealed that the
orientational order survives in a wide range of lateral
interactions.

A notable feature is that nematic order is only stable for
sufficiently large aspect ratios while systems of short rods do
not show nematic order at all. The long-range orientational order
also disappears in the case of irreversible adsorption (as is the
case of papers I and II), where the distribution of adsorbed
objects is different from that obtained at equilibrium. Thus, the
irreversible adsorption leads to intermediate states characterized
by an isotropic distribution of the directions of the adsorbed
molecules, and to a final state (known as jamming state), in which
no more objects can be deposited due to the absence of free space
of appropriate size and shape (the jamming state has infinite
memory of the process and the orientational order is purely local)
\cite{EVANS}.

It is of interest then to study the percolation problem of
$k$-mers at equilibrium, not only because of this problem has not
been treated in the literature, but also because, depending on the
values of the density and the size $k$, the model leads either to
an isotropic system (as studied in paper I) or an anisotropic
system (as studied in paper II). In this sense, the aim of the
present work is to study, by Monte Carlo (MC) simulations and
finite-size scaling analysis, the effect of the orientational
order on the percolation properties for a system of rigid rods
adsorbed at equilibrium on two-dimensional square lattices.

\section{Model and Monte Carlo simulation}\label{model}

The calculations have been developed for straight rigid rods of
length $k$, with $k$ ranging between 2 and 12. Small adsorbates
with spherical symmetry would correspond to the monomer limit ($k
=1$). The distance between $k$-mer units is assumed to be equal to
the lattice constant; hence exactly $k$ sites are occupied by a
$k$-mer when adsorbed. The surface was represented as a 2D square
lattice of $M=L \times L$ adsorptive sites with periodic boundary
conditions. The only interaction between different rods is
hard-core exclusion: no site can be occupied by more than one
$k$-mer.

To describe a system of $N$ $k$-mers adsorbed on $M$
sites at a given temperature $T$ and chemical potential $\mu$, the
occupation variable $c_j$ was introduced ($c_j=0$ or 1, if the
site $j$ is empty or occupied by a $k$-mer unit, respectively).
Then the adsorbed phase is characterized by the Hamiltonian
$H=\left(\epsilon_0-\mu \right) \sum_j c_j$, where the sum run
over the $M$ sites and $\epsilon_0$ is the adsorption energy of a
$k$-mer unit (in the simulations, $\epsilon_0$ is set equal to
zero without any loss of generality).

The thermodynamic equilibrium is reached in the grand canonical
ensemble by using the hyper-parallel tempering Monte Carlo (HPTMC)
simulation method \cite{Yan,Huku} combined with an efficient
cluster algorithm (the pocket algorithm) \cite{Krauth}.

The HPTMC method consists in generating a compound system of $R$
noninteracting replicas of the system under study. The $i$-th
replica is associated with a chemical potential $\mu_{i}$. To
determine the set of chemical potentials, $\{\mu_i \}$, we set the
lowest chemical potential, $\mu_1$, in the isotropic phase where
relaxation (correlation) time is expected to be very short and
there exists only one minimum in the free energy space. On the
other hand, the highest chemical potential, $\mu_R$, is set in the
nematic phase whose properties we are interested in. Finally, the
difference between two consecutive chemical potentials, $\mu_i$
and $\mu_{i+1}$ with $\mu_i < \mu_{i+1}$, is set as $\Delta \mu =
\left(\mu_{1} - \mu_R \right)/(R-1)$ (equally spaced chemical
potentials). The parameters used in the present study were as
follows: $R=40$, $\mu_1=-2.5$ and $\mu_R=2.375$. With these values
of the chemical potential, the corresponding values of the surface
coverage varied from $\theta_1(\mu_1) \approx 0.21 $ to
$\theta_R(\mu_R)\approx 0.88$ for $k=2$, and from
$\theta_1(\mu_1)\approx 0.33$ to $\theta_R(\mu_R)\approx 0.81$ for
$k=12$.

Under these conditions, the algorithm to carry out the simulation
process is built on the basis of two major subroutines: {\it
replica-update} and {\it replica-exchange}.

\noindent {\it Replica-update}: The adsorption-desorption and
pocket procedure is as follows: (1) One out of $R$ replicas is
randomly selected. (2) A linear $k$-uple of nearest-neighbor
sites, belonging to the replica selected in (1), is chosen at
random. Then, if the $k$ sites are empty, an attempt is made to
deposit a rod with probability $W = \min \{1, \exp(\beta \mu)\}$;
if the $k$ sites are occupied by units belonging to the same
$k$-mer, an attempt is made to desorb this $k$-mer with
probability $W = \min \{1, \exp(-\beta \mu)\}$; and otherwise, the
attempt is rejected. In addition, the pocket algorithm
\cite{Krauth} is applied in order to reach equilibrium in a
reasonable time. This algorithm includes the following steps: (1)
Choose randomly a lattice symmetry axis and a seed $k$-mer.
The seed $k$-mer now is the sole element of a set $\mathcal{P}$ and the rest
of $k$-mers belongs to a set $\mathcal{O}$. (2) Move elements from
the set $\mathcal{P}$ to $\mathcal{O}$ in the following way: Pick
up an arbitrary element $i$ of $\mathcal{P}$ and reflect with
respect to the symmetry axis. If $i$ overlaps with other elements
of $\mathcal{O}$, the latter are transferred from $\mathcal{O}$ to
$\mathcal{P}$. (3) Repeat the step (2) until the set
$\mathcal{P} = \varnothing$.

\noindent {\it Replica-exchange}: Exchange of two configurations
$X_i$ and $X_j$, corresponding to the $i$-th and $j$-th replicas,
respectively, is tried and accepted with probability $W=\min \{ 1,
\exp{(-\Delta)}\}$. Where $\Delta$ in a nonthermal grand canonical
ensemble is given by $[-\beta(\mu_{j}-\mu_{i}) \,(N_{j}-N_{i})]$,
and $N_{i}$ ($N_{j}$) represents the number of particles of the
$i$-th ($j$-th) replica.

The complete simulation procedure is the following: (1)
replica-update, (2) replica-exchange, and (3) repeat from step (1)
$RM$ times. This is the elementary step in the simulation process
or Monte Carlo step (MCs).

For each value of the chemical potential $\mu_i$, the equilibrium
state can be well reproduced after discarding the first $r_0$ MCs.
Then, a set of $m$ samples in thermal equilibrium is generated.
The corresponding surface coverage $\theta_i(\mu_i)$ is obtained
through simple averages over the $m$ samples ($m$ MCs).

As mentioned before in Ref. \cite{GHOSH}, the relaxation time
increases very quickly as the $k$-mer size increases.
Consequently, MC simulations for large adsorbates are very time
consuming and may produce artifacts related to non-accurate
equilibrium states. In order to discard this possibility,
equilibration times $r_0$ of the order O($10^6$ MCs) were used in
this study, with an effort reaching almost the limits of our
computational capabilities \cite{foot2}.

\section{Finite-size scaling results}

The central idea of the pure percolation theory is based in
finding the minimum concentration of elements (sites or bonds) for
which a cluster extends from one side to the opposite one of the
system. At this particular value of the concentration, the
percolation threshold $\theta_c$, a second-order phase transition
occurs in the system, which is characterized by well-defined
critical exponents \cite{Stauffer}.

The finite-size scaling theory \cite{Binder} gives us the basis to
achieve the percolation threshold and the critical exponents of a
system with a reasonable accuracy. For this purpose, the
probability $R=R^X _{L,k}(\theta)$ that a $L \times L$ lattice
percolates at a concentration $p$ of sites occupied by rods of
size $k$ can be defined \cite{Stauffer,Yone}. Here, the following
definitions can be given according to the meaning of $X$: a)
$R^{R(D)}_{L,k}(\theta)=$ the probability of finding a rightward
(downward) percolating cluster; and b) $R^{U}_{L,k}(\theta)=$ the
probability of finding either a rightward {\bf or} a downward
percolating cluster.

In the Monte Carlo (MC) simulations, each run consists of the
following steps: (a) the construction of $m=2 \times 10^5$ samples
for the desired fraction $\theta(\mu)$ of occupied sites (according to the
HPTMC method described above); and (b) the cluster analysis by
using the Hoshen and Kopelman algorithm \cite{Hoshen}. In the last
step, the existence of a percolating island is verified for each
sample. The spanning cluster could be determined by using the
criteria $R$, $D$ and $U$. This scheme is carried out for
obtaining the number $m^X$ of samples for which a percolating
cluster of the desired criterion $X$ is found. Then,
$R^{X}_{L,k}(\theta)=m^X/m$ is defined and the procedure is
repeated for different values of $\theta$, $L$ and $k$.

Besides the parameter $R$, the percolation order parameter $P =
\langle S_L \rangle/L^2$ \cite{Biswas,Chandra} has been measured,
where $S_L$ represents the size of the largest cluster and
$\langle ... \rangle$ means the average over the $m$ samples. The
corresponding percolation susceptibility $\chi$ has also been
calculated, $\chi=\left[ \langle S_L^2 \rangle - {\langle S_L
\rangle}^2 \right]/L^2$.

In Figure 1, the probability $R^{U}_{L,k}(\theta)$ is presented  for
$k=3$ (left), $k=5$ (middle) and $k=7$ (right) and different
lattice sizes $L/k =5,10,15,20$ \cite{foot1}, as indicated in the
caption of the figure. Several conclusions can be drawn from the
figure. First, curves cross each other in a unique point
$R^{X^*}_k$ (measured in the vertical axes), which depends on the
criterion $X$ used (data corresponding to criteria $R$ and $D$ are
not shown in the figure for clarity) and those points are located
at very well defined values in the $\theta$-axes determining the
critical percolation threshold (measured in the horizontal axes)
for each $k$. The values of $\theta_c(k)$ obtained by following
the procedure of Figure 1 are collected in Table 1 and are plotted
in Figure 2a (full circles). A nonmonotonic size dependence is
observed for the percolation threshold, which decreases for small
particles sizes, goes through a minimum around $k=5$, and finally
asymptotically converges towards a definite value for large
segments. This striking behavior can be interpreted as a
consequence of the IN phase transition occurring in the system for
large values of $k$ \cite{GHOSH,EPL1,PHYSA19,JCP7,JSTAT1}. To
understand this effect, it is convenient first to recall some
results about percolation of straight rigid rods isotropically
deposited (or isotropic $k$-mers) \cite{EJPB1} and straight rigid
rods deposited along one of the directions of the lattice (or
aligned $k$-mers) \cite{LONGO}.

In the first case (isotropic $k$-mers) \cite{EJPB1}, the rods are
deposited randomly and irreversibly. The process is as follows: a
linear $k$-uple of nearest-neighbor sites is randomly selected; if
it is vacant, a $k$-mer is then deposited on those sites.
Otherwise, the attempt is rejected. This filling process, known as
Random Sequential Adsorption model, leads to states characterized
by an isotropic distribution of the directions of the adsorbed
$k$-mers. In the case of aligned $k$-mers \cite{LONGO}, the
adsorption process is also irreversible, but this time the
$k$-mers are deposited along one of the directions of the lattice,
forming a nematic phase.

As it can be observed from Figure 2a, the percolation threshold
curves corresponding to isotropic (open circles) and aligned (open
squares) $k$-mers decrease upon increasing $k$. At the beginning,
for small values of $k$, the curves rapidly decrease. However,
they flatten out for larger values of $k$ and finally
asymptotically converge towards a definite value $\theta^*_c$ (as
$k \rightarrow \infty$), being $\theta^*_c \approx 0.46$ for
isotropic $k$-mers and $\theta^*_c \approx 0.54$ for aligned
$k$-mers.

Now, the nonmonotonic behavior observed in Figure 2a can be
explained on the basis of the results for isotropic and aligned
$k$-mers. In fact, for small values of $k$, there is no
orientational order in the equilibrium adlayer and $\theta_c(k)$
decreases following the curve corresponding to isotropic $k$-mers.
On the other hand, for large values of $k$, a nematic phase,
characterized by a big domain of parallel $k$-mers, is formed on
the surface at an intermediate density $\theta_{IN}$
\cite{GHOSH,EPL1,PHYSA19,JCP7,JSTAT1}. The values of $\theta_{IN}$
reported in Ref. \cite{JCP7} are listed in the third column of
Table 1. For $k \geq 9$, $\theta_c > \theta_{IN}$ and,
consequently, (1) the percolation properties of the orientational
phase are similar to the ones corresponding to perfectly aligned
$k$-mers and (2) $\theta_c(k)$ follows the curve corresponding to
aligned $k$-mers. As a consequence of the behavior of
$\theta_c(k)$ in the limits of small and large $k$-mers, a marked
minimum is observed at intermediate values of $k$.

Typical equilibrium states are shown schematically in parts
(b)-(d) of Figure 2. In the case of Figure 2b [point $i$ in Figure
2a: $k=2$ and $\theta \approx 0.53$], the equilibrium state
corresponds to an isotropic and nonpercolating state. The
situation is different in Figure 2c [point $f$ in Figure 2a: $k=5$
and $\theta \approx 0.53$], where the equilibrium state is an
isotropic and percolating state. Thin bars represent isolated
$k$-mers and thick bars (red bars in the online version)
correspond to $k$-mers belonging to the percolating cluster.
Finally, a nematic and nonpercolating state is observed in Figure
2d [point $j$ in Figure 2a: $k=10$ and $\theta \approx 0.53$].

A second conclusion of Figure 1 is that the crossing points of the percolation cumulants
$R^{U^*}_k$ (as $R^{R^*}_k$ and $R^{D^*}_k$) vary as the $k$-mer
size is increased. This behavior was already observed
in the case of perfectly aligned $k$-mers irreversibly deposited
on 2D lattices \cite{LONGO} and can be attributed to the strong
anisotropy introduced in the system by the IN phase transition
occurring for large $k$-mers. In the case of thermal transitions,
the link between anisotropy and a nonuniversal behavior of the
intersection point of the cumulants has been discussed in
\cite{Selke1,Selke2}. In fact, as pointed out by Selke et al.
\cite{Selke1,Selke2}, the measure of the cumulant intersection may
depend on various details of the model, which do not affect the
universality class, in particular, the boundary condition, the
shape of the lattice, and the anisotropy of the system. In order
to confirm or discard this hypothesis, the critical exponents
$\nu$, $\beta$ and $\gamma$ were calculated as follows.

The value of $\nu$ can be obtained through the scaling
relationship for $R^X_{L,k}(\theta)$
\begin{equation}
R^X_{L,k}(\theta)=\overline{R^X_k}\left[ \left( \theta - \theta_c
\right) L^{1/\nu} \right], \label{functionR}
\end{equation}
being $\overline{R^X_k}(u)$ the scaling function and $u \equiv
\left( \theta - \theta_c \right) L^{1/\nu}$. Then, the maximum of
the derivative of Eq. (\ref{functionR}) leads to
$\left(dR^X_{L,k}/d\theta \right)_{\rm max} \propto L^{1/\nu}$.

In the inset of Figure 3, $\left(dR^X_{L,k}/d\theta \right)_{\rm
max}$ has been plotted as a function of $L/k$ (note the log-log
scale) for different $k$-mers as indicated. According to Eq.
(\ref{functionR}) the slope of each line corresponds to $1/ \nu$.
As it can be observed, the slopes of the curves remain constant
(and close to 3/4) for all values of $k$. The results coincide,
within numerical errors, with the exact value of the critical
exponent of the ordinary percolation $\nu=4/3$.

The scaling behavior can be further tested by plotting
$R^{X}_{L,k}(\theta)$ vs $\left(\theta - \theta_c \right)
L^{1/\nu}$ and looking for data collapsing. Using the values of
$\theta_c$ previously calculated and the exact value $\nu=4/3$, an
excellent scaling collapse was obtained (Figure 3) for all value of
$k$-mer size. This leads to independent control and consistency
check of the numerical value of the critical exponent $\nu$.

The critical exponents $\beta$ and $\gamma$ were obtained from the
scaling behavior of $P$ and $\chi$ \cite{Stauffer},
\begin{equation}
P=L^{-\beta/\nu} \overline{P}\left[ | \theta - \theta_c |
L^{1/\nu} \right], \label{functionP}
\end{equation}
and
\begin{equation}
\chi=L^{\gamma/\nu} \overline{\chi}\left[ \left( \theta - \theta_c
\right) L^{1/\nu} \right], \label{functionchi}
\end{equation}
where $\overline{P}$ and $\overline{\chi}$ are scaling functions
for the respective quantities. According to Eqs. (\ref{functionP})
and (\ref{functionchi}), Figure 4 shows the collapse of the curves
of $P$ and $\chi$ (inset) for a typical $k$-mer size ($k=5$) and
different lattice sizes as indicated. The study was repeated for all studied values of $k$ (these data are not shown here for lack of space). The data scaled extremely
well using the exact percolation exponents $\beta=5/36$ and
$\gamma=43/18$. The results obtained in Figures 3 and 4 clearly
supports that, as in the case of isotropic rods \cite{EJPB1}, the
universality class of the problem does not depend on the $k$-mer
size.

\section{Conclusions}

In this work, Monte Carlo simulations and finite-size scaling theory
have been used to study the percolation properties of straight
rigid rods of length $k$ adsorbed at equilibrium on
two-dimensional square lattices.

A nonmonotonic size dependence
was found for the percolation threshold $\theta_c$, which
decreases for small particles sizes [following the curve of
$\theta_c(k)$ for $k$-mers irreversibly and isotropically
deposited], goes through a minimum, and finally asymptotically
converges towards a definite value for large segments [following
the curve of $\theta_c(k)$ for aligned $k$-mers irreversibly
deposited along one of the directions of the lattice]. This
striking behavior has been interpreted as a consequence of the IN
phase transition occurring in the system for large values of $k$.
Finally, the analysis of the critical exponents $\nu$, $\beta$ and
$\gamma$ revealed that the percolation phase transition involved
in the problem considered in the present paper belongs to the same
universality class of the ordinary random percolation.

\subsection*{Acknowledgment}
This work was supported in part by CONICET (Argentina) under
project number PIP 112-200801-01332; Universidad Nacional de San
Luis (Argentina) under project 322000 and the National Agency of
Scientific and Technological Promotion (Argentina) under project
PICT-2010-1466.

\newpage\parindent 0 cm \parskip=5mm

\section*{Table and Figure Captions}

\noindent Table 1: Values of the percolation threshold $\theta_c$
for $k$ ranging from $1$ to $12$ and of the isotropic-nematic
critical density $\theta_{IN}$ \cite{JCP7} for $k$ between $7$ to
$12$. Note that the IN phase transition on a square lattice with
two allowed orientations occurs only for $k \geq 7$
\cite{GHOSH,EPL1,PHYSA19,JCP7,JSTAT1}. In the case of $k=1$, the
problem of percolation of monomers on square lattices has been one
of the most studied percolation models in the literature, and the
percolation threshold has been measured multiple times as
$0.592746xx$, where the last two decimal places (and error) are
$21(13)$ \cite{Newman}, $21(33)$ \cite{Oliveira}, $03(09)$
\cite{Lee1}, and $06(05)$ \cite{Feng}. For a complete overview on
this topic we refer the reader to Ref. \cite{Lee2}.

\noindent Fig. 1: Fraction of percolating lattices $R^U_{L,k}$ as
a function of the concentration $\theta$ for $k=3$ (left), $k=5$
(middle) and $k=7$ (right) and different lattice sizes: $L/k=5$,
diamonds; $L/k=10$, squares; $L/k=15$, circles and $L/k=20$,
triangles. Horizontal dashed lines show the intersection points
$R^{X^*}_k$. Vertical dashed line denotes the percolation
threshold $\theta_c$ in the thermodynamic limit.

\noindent Fig. 2: (Color online) (a) The percolation threshold
$\theta_c$ as a function of $k$ for straight rigid rods on 2D
square lattices and three different adsorption processes: nematic
irreversible adsorption (open squares), isotropic irreversible
adsorption (open circles) and equilibrium adsorption (solid
circles). The dashed lines are simply a guide for the eye. In all
cases, the error bar is smaller that the size of the symbols. The
asymptotic limits, $\theta^*_c$, [$\theta_c(k)$ for $k \rightarrow
\infty $] are shown. (b) Schematic representation of the
equilibrium state for $k=2$ and $\theta \approx 0.53$ (point $i$
in the figure). (c) Same as (b) for $k=5$ and $\theta \approx
0.53$ (point $f$ in the figure). Thin bars represent isolated
$k$-mers and thick bars (red bars in the online version)
correspond to $k$-mers belonging to the percolating cluster. (d)
Same as (b) for $k=10$ and $\theta \approx 0.53$ (point $j$ in the
figure).

\noindent Fig. 3: Data collapsing of the fraction of percolating
samples $R^{U}_{L,k}(\theta)$ as a function of the argument
$\left(\theta - \theta_c \right) L^{1/\nu}$. Each set of curves
corresponds to a different value of $k$ as indicated. For each
$k$, different lattice sizes ($L/k$ = 5, 10, 15 and 20) have been
considered. Inset: $\ln\left(\frac{dR^U_{L,k}}{d\theta}
\right)_{\rm max}$ as a function of $\ln(L/k)$ for different
values of $k$ as indicated. According to Eq. (\ref{functionR}) the
slope of each line corresponds to $1/\nu$.

\noindent Fig. 4: Data collapsing of the order parameter, $P
L^{\beta/\nu}$ vs $|\theta - \theta_c| L^{1/\nu}$, and of the
susceptibility, $\chi L^{-\gamma/\nu}$ vs $\left(\theta - \theta_c
\right) L^{1/\nu}$ (inset) for $k=5$. The plots were made using
the exact percolation exponents $\nu=4/3$, $\beta=5/36$ and
$\gamma=43/18$.

\newpage

\begin{center}
Table 1
\end{center}

$$
\begin{array}{|c|c|c|}
\hline  \hline
{\rm {\it k}-mer \ size}, k & {\rm percolation \ threshold}, \theta_c & {\rm isotropic-nematic \ critical \ density}, \theta_{IN}\\
\hline
1 & 0.592746xx(yy) & --- \\
\hline
2 & 0.586(1) & --- \\
\hline
3 & 0.575(1) & --- \\
\hline
4 & 0.567(1) & --- \\
\hline
5 & 0.563(1) & --- \\
\hline
6 & 0.559(1) & --- \\
\hline
7 & 0.593(1) & 0.729(3) \\
\hline
8 & 0.586(1) & 0.648(3) \\
\hline
9 & 0.575(1) & 0.569(3) \\
\hline
10 & 0.567(1) & 0.502(3) \\
\hline
11 & 0.563(1) & 0.457(5) \\
\hline
12 & 0.559(1) & 0.413(5) \\
\hline
\end{array}
$$

\begin{center}

\begin{figure}
\includegraphics[scale=0.8]{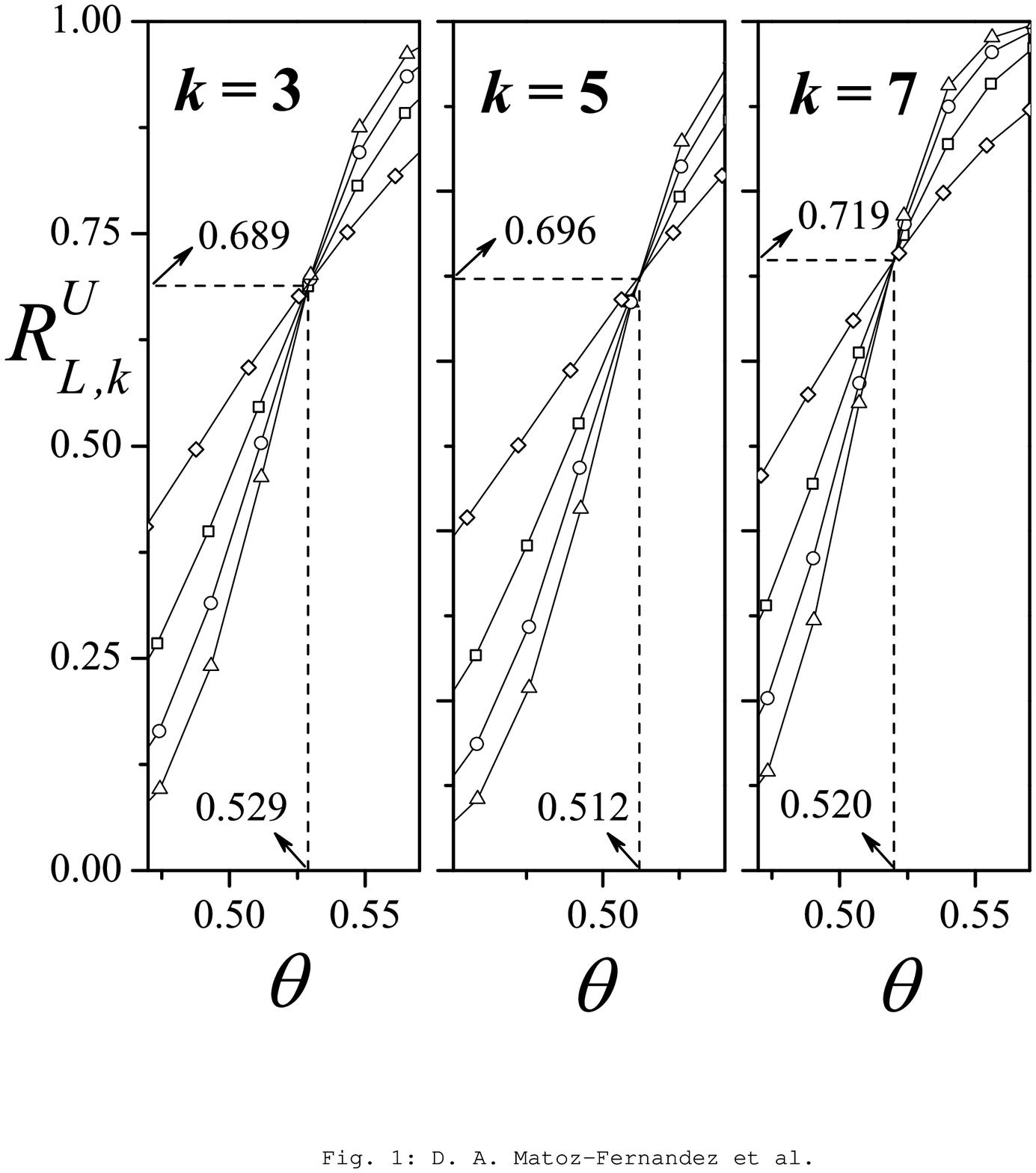}
\end{figure}

\begin{figure}
\includegraphics[scale=0.8]{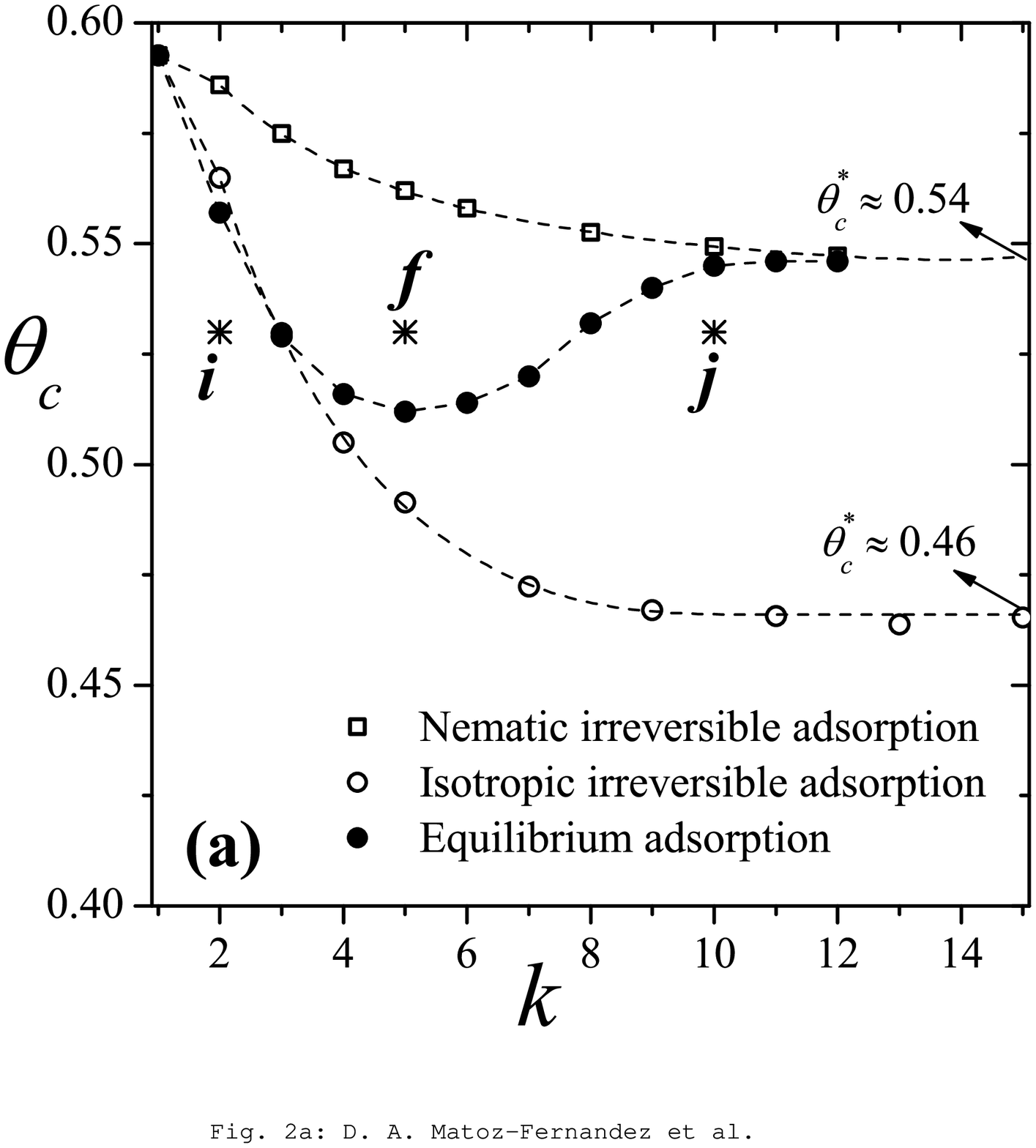}
\end{figure}

\begin{figure}
\includegraphics[scale=0.8]{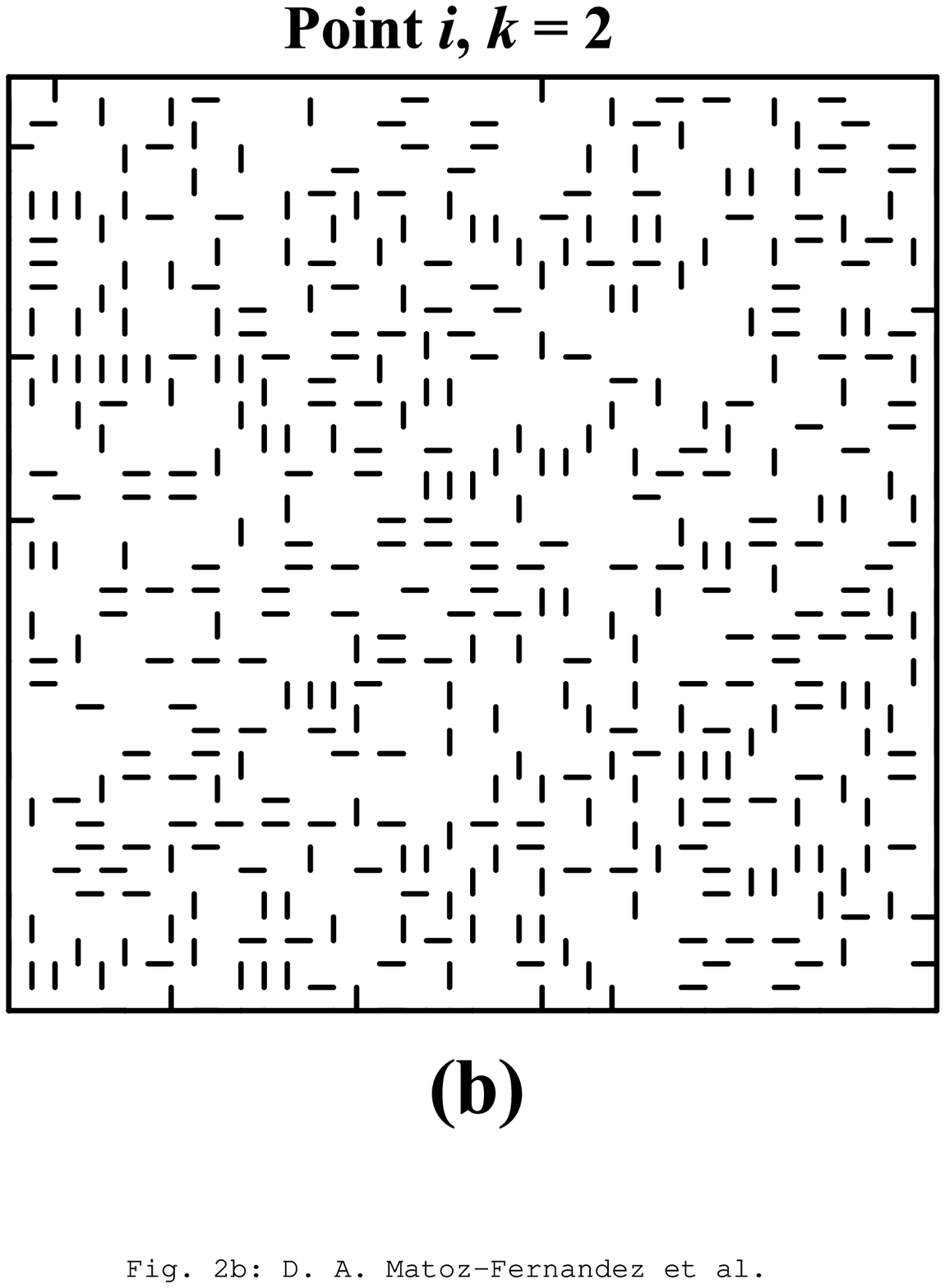}
\end{figure}

\begin{figure}
\includegraphics[scale=0.8]{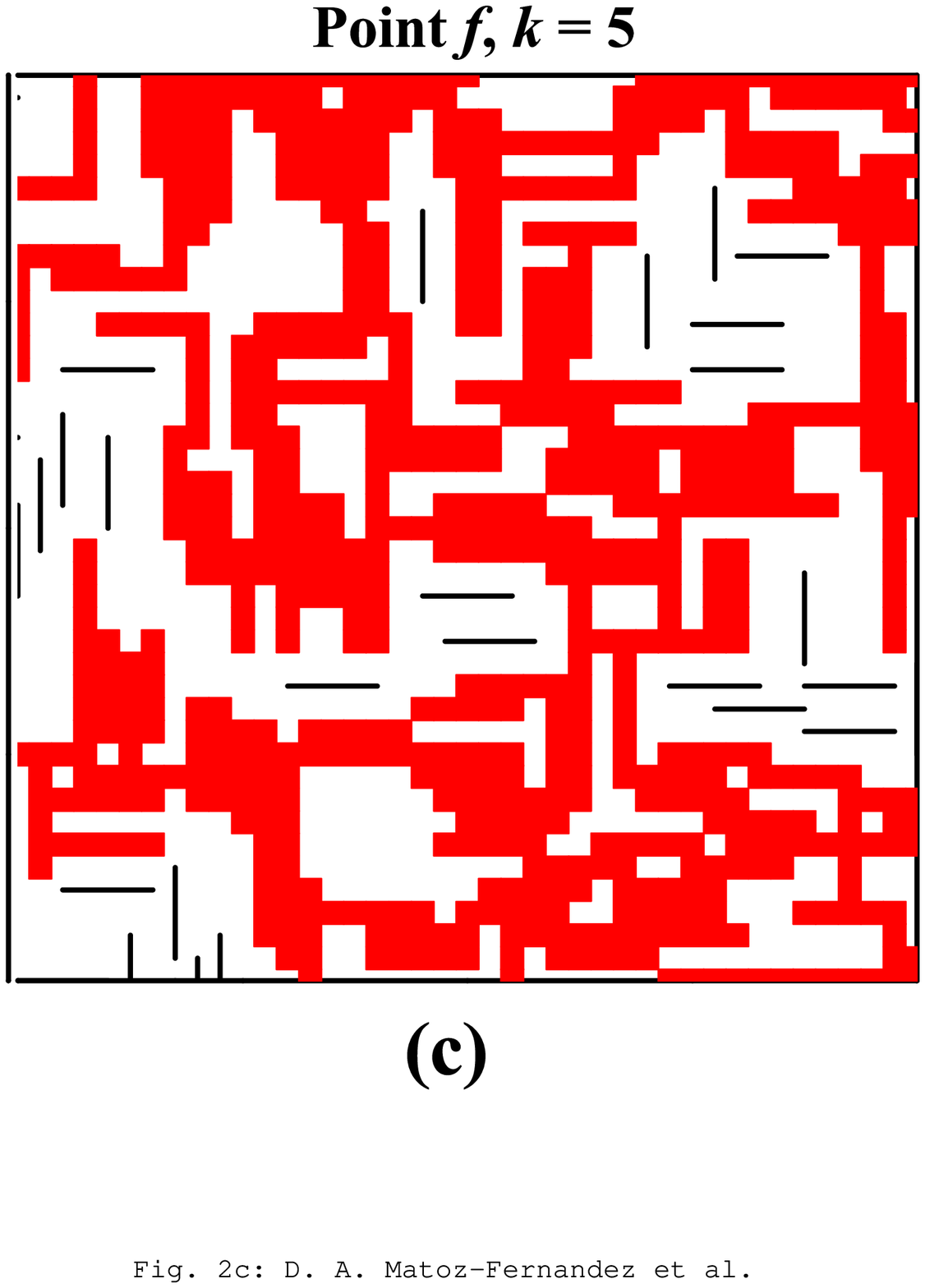}
\end{figure}

\begin{figure}
\includegraphics[scale=0.8]{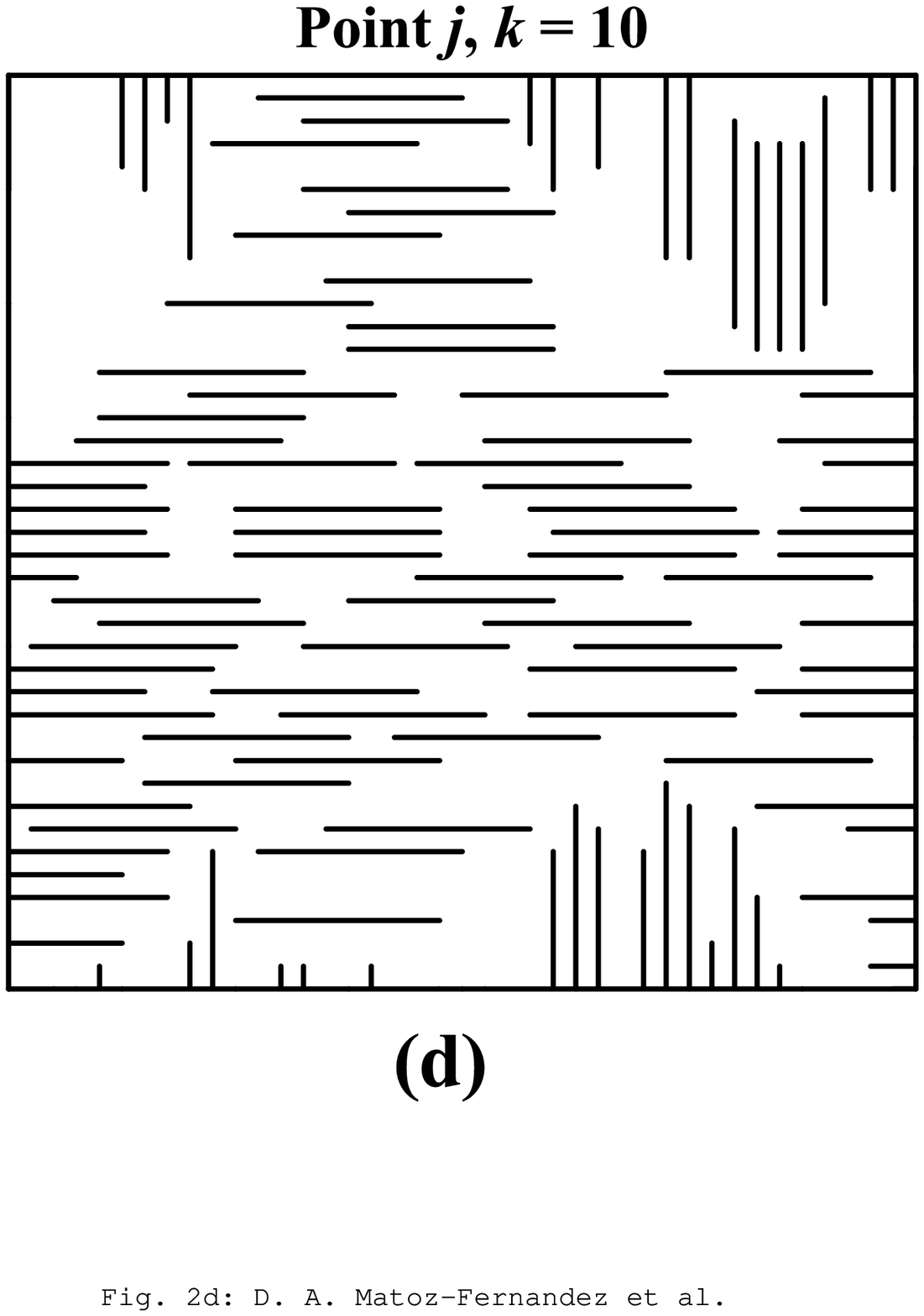}
\end{figure}

\begin{figure}
\includegraphics[scale=0.8]{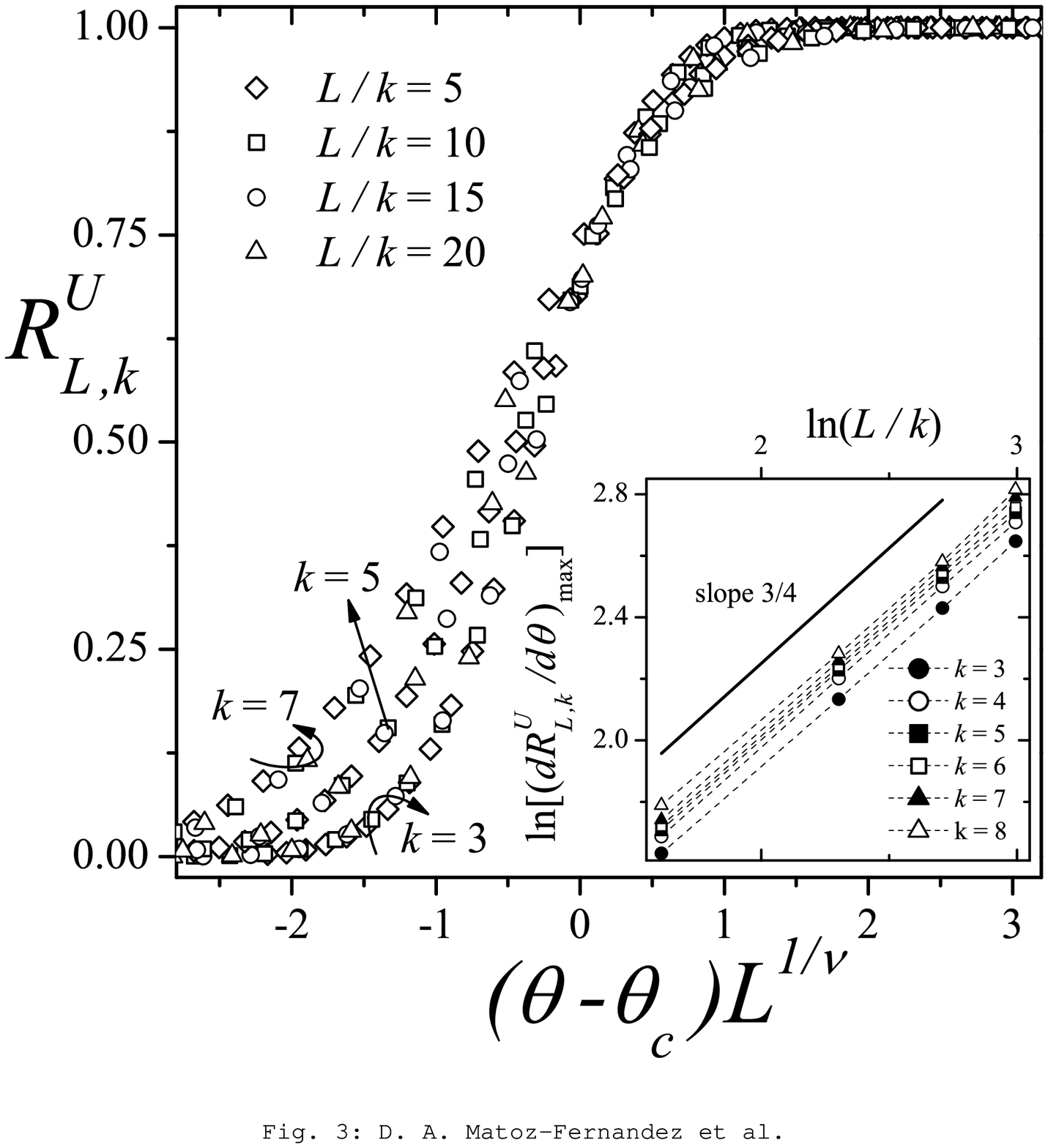}
\end{figure}

\begin{figure}
\includegraphics[scale=0.8]{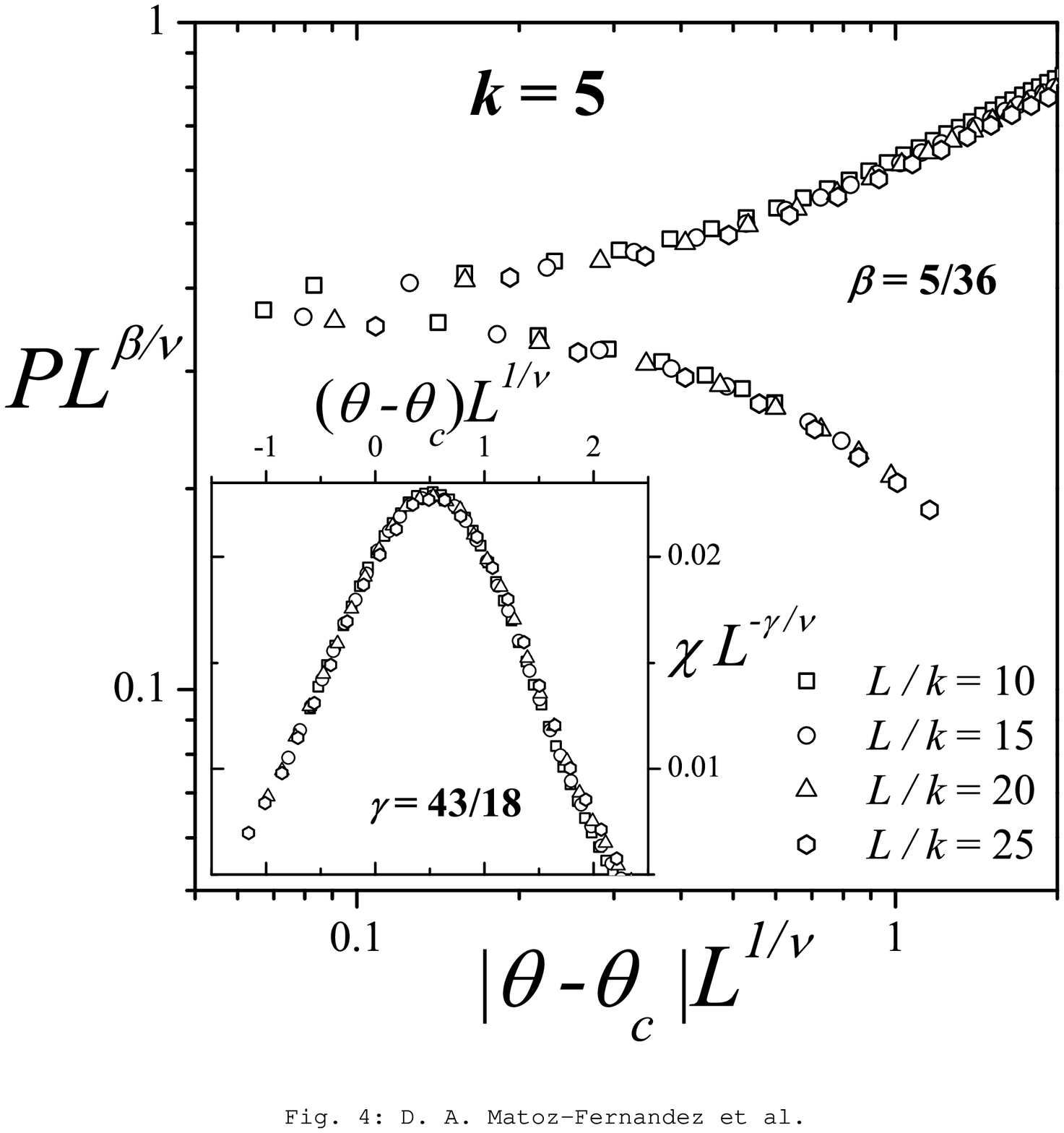}
\end{figure}

\end{center}


\begin{thebibliography}{10}
\bibitem{foot2} As an example, it took about 90 days to
obtain the adsorption isotherm corresponding to $k = 12$ and
$L/k=20$ on one node of the BACO parallel cluster. This facility,
located at Instituto de Física Aplicada, Universidad Nacional de
San Luis-CONICET, San Luis, Argentina, consists of 50 CPUs each
with an Intel Core i7 processor running at 2.93 GHz and 512MB of
RAM per core.
\bibitem{foot1} As in previous work \cite{EJPB1}, the ratio $L/k$ is kept constant for
avoiding spurious effects due to the $k$-mer size in comparison
with the lattice size.
\end{thebibliography}
\end{document}